# *DXRD*, a user-friendly suite of two- and multiple-beam dynamical X-ray diffraction programs

XianRong Huang* and Lahsen Assoufid

Advanced Photon Source, Argonne National Laboratory, 9700 South Cass Avenue, Lemont, IL 60439, USA. *Corresponding email: xiahuang@anl.gov

**Abstract**

The *DXRD* program suite consisting of a series of dynamical-theory programs is introduced for computing dynamical X-ray diffraction from single crystals. Its interactive graphic user interfaces (GUIs) allow general users to make complicated calculations with minimal effort. It can calculate plane-wave Darwin curves of single crystals (or multiple crystals) for both the Bragg and Laue cases, including grazing-incidence diffraction and backward diffraction (with Bragg angles approaching 90°). It is also capable of simulating rocking curves for divergent incident X-ray beams with finite bandwidths. A unique capability of *DXRD* is that it provides, for the first time, a convenient GUI-based multiple-beam diffraction program that can accurately compute arbitrary $N$-beam diffraction of any geometry using a universal $4N \times 4N$ matrix method. *DXRD* also provides a mapping program for plotting the entire multiple-beam diffraction lines (monochromator glitches) in the azimuth-energy coordinate system. All these functions make *DXRD* a convenient and powerful software tool for designing crystal-based synchrotron/X-ray optics (monochromators, analyzers, polarizers, phase plates, etc) and for crystal characterization, X-ray spectroscopy and X-ray diffraction teaching.

**Synopsis:** The *D*XRD program suite consisting of a series of dynamical-theory programs (particularly the rigorous multiple-beam diffraction programs) with friendly interactive graphic user interfaces is introduced for computing arbitrary 2- and *N*-beam X-ray diffraction from single crystals.

# 1. Introduction

The Dynamical theory of X-ray diffraction is a well-established wave theory that can accurately calculate X-ray diffraction from (nearly) perfect single crystals [1,2]. It has wide applications in crystal characterization and in the designing of crystal-based synchrotron/X-ray optics. The theory is conceptually simple, but practical computation of crystal diffraction using the dynamical theory is, in fact, a daunting task even for experts since it requires complicated structure factor calculations for arbitrary crystals and X-ray energies, consideration of various geometrical configurations, convolution with the beam conditions and instrument functions, etc. Thus, most researchers rely on limited computer programs to perform dynamical theory computations.

Among the freely available programs, the *XOP* (X-ray Oriented Programs) package has been the main software for dynamical-theory computation (Sánchez del Río & Dejus, 2011) and is widely used by scientists at synchrotron light sources. In parallel, Stepanov (2021) has been maintaining the web-based *X-ray Server* (https://x-server.gmca.aps.anl.gov) that also has a collection of popular dynamical diffraction programs. To run these programs, however, the users need in-depth knowledge of the dynamical theory with its conventions used, which often requires extensive study of documentations and textbooks. Furthermore, some special diffraction conditions, such as non-coplanar diffraction, backward diffraction, multiple-beam diffraction (MBD) and convolution with instrument functions, may not be included in these programs or are only partially treated without universal and reliable algorithms.

Here we introduce another package, called *DXRD*, that consists of a series of dynamical-theory programs integrated in a single executable file (see Fig. 1). Distinctively, *DXRD* has interactive graphic user interfaces (GUIs) that can guide general users (including non-experts with little dynamical theory knowledge) to make complicated calculations *only with a few mouse clicks* without extensive consultations with documentation. It can calculate plane-wave rocking curves for both the Bragg and Laue cases, and it is also capable of simulating rocking curves for real incident X-ray beams with finite bandwidths and angular divergence. Another unique capability of *DXRD* is that it provides, for the first time, a convenient GUI-based MBD program that can rapidly compute $N$-beam diffraction (including coplanar or non-coplanar two-beam diffraction). This program is based on a full $4N \times 4N$ matrix method that is valid for any diffraction geometry. Its computing algorithms have been highly optimized so that computing a two-dimensional (2D) MBD intensity contour with ~10,000 points can be finished with minimal delays.

*DXRD* was written with Visual C++ and the software package is available at GitHub repository https://github.com/xhuang555-anl/DXRD_Release. It is free for academic research. The entire *DXRD* software is a stand-alone application that can run on any Windows computers or laptops without the necessity of installation and does not depend on external resources. We hope that it will be a convenient software tool for researchers in the communities of synchrotron/X-ray optics, crystallography, X-ray spectroscopy, etc to perform fast development of synchrotron components or make interpretation of experimental data easier. It can also be a teaching tool for students to learn X-ray diffraction.

## 2. General features and capabilities of *DXRD*

Similar to the sister program *LauePt* (Huang, 2010), *DXRD* does not require the user to register crystal structures to a central database. Instead, it allows the user to compile a simple text file for any crystal structure. The file only requires the information of (1) a crystal name, (2) the lattice constants $a, b, c, \alpha, \beta, \gamma$, (3) the number of atoms in the unit cell, (4) the positions of the atoms in the unit cell, and (5) the Debye characteristic temperatures. Then DXRD can import the file and calculate the structure factors of any reflections at any X-ray energies using either the tabulated-coefficient method (similar to *XOP*) or the Brennan-Cowan method (Brennan & Cowan, 1992). The user can also manually input structure factors.

Based on the structure factors, *DXRD* can quickly calculate the plane-wave rocking curves for both the Bragg-reflection geometry and the Laue-transmission geometry, including asymmetric reflection, grazing-incidence/exit geometry, backward reflection (with the Bragg angle $\theta_B \rightarrow 90°$), etc. The diffraction curve can also be calculated as a function of the photon energy $E$, from which one obtains the spectral bandpass of the crystal as an X-ray monochromator or analyzer. The programs provide abundant flexibility for the user to freely change any of the parameters, followed by instant update of the diffraction curves. These features are particularly useful in the designing of crystal-based X-ray optics.

In practical experiments, the incident X-ray beam always has angular divergence and a spectral bandwidth, which makes the measured rocking curve much wider than the intrinsic Darwin curve in most cases. The "Double-Crystal Diffraction" program of *DXRD* is for calculating the convoluted rocking curves based on an X-ray beam with finite divergence and energy spread that are conditioned by an upstream monochromator. When the monochromator is set to be a double-crystal monochromator (DCM), which is the typical setup at most synchrotron beamlines, the program can simulate the real rocking curves measured using synchrotron beams. It can also simulate the double-crystal rocking curves from lab-source-based diffractometers, where the incident beam has large divergence.

The "Multiple-Crystal Monochromator" program is for calculating and optimizing the angular acceptance, bandwidths and efficiency of monochromators consisting of multiple crystals in arbitrary arrangements. It is still in the early stages of development. Currently it can be used to model DCMs, channel-cut monochromators, four-bounce Bartels monochromators (Bartels, 1983; Loxley *et al.*, 1995), etc.

## 3. Computation of MBD and mapping of MBD lines (monochromator glitches)

As mentioned above, a unique capability of *DXRD* is that it provides a rigorous GUI-based MBD program that can compute MBD (Colella, 1974; Stetsko & Chang, 1997; Chang, 2004) of any geometry with a universal algorithm. It is based on the *Fourier coupled-wave diffraction theory* (FCWDT) (Huang *et al.*, 2013) that was originally developed for modelling optical diffraction from photonic crystals with thousands of diffraction orders (Huang *et al.*, 2010). This algorithm is fast for MBD because X-ray diffraction can only involve up to tens of reflections.

To compute MBD, the user first chooses a *primary reflection* **G** and a *horizontal reference direction* (HRD) perpendicular to **G** in reciprocal space (see Fig. 1). Then the program establishes a Cartesian coordinate system (CS) with the **Z** and **X** axes parallel to **G** and the HRD, respectively, as shown in Fig. 2(*a*). The direction of the incident beam is described by the *diffraction angle* $\theta$ (relative to the diffracting planes) and the *azimuth angle* $\Phi$ (relative to the **X**-axis). Thus, in Fig. 2(*a*) the incident wavevector is

$$\mathbf{K}_0 = K(\cos\Phi\cos\theta\,\hat{\mathbf{x}} + \sin\Phi\cos\theta\,\hat{\mathbf{y}} - \sin\theta\,\hat{\mathbf{z}}), \tag{1}$$

and the diffracted wavevector of the primary reflection **G** is

$$\mathbf{K}_G \approx K(\cos\Phi\cos\theta\,\hat{\mathbf{x}} + \sin\Phi\cos\theta\,\hat{\mathbf{y}} + \sin\theta\,\hat{\mathbf{z}}), \tag{2}$$

where $\hat{\mathbf{x}}, \hat{\mathbf{y}}, \hat{\mathbf{z}}$ are unit vectors along the **X**, **Y**, **Z** axes, respectively, and $K = 1/\lambda$ ($\lambda$ the X-ray wavelength). In general, the crystal surface has an *offcut angle* with respect to the diffracting planes of **G**, denoted by $\Omega_{\text{off}}$ in Fig. 2(*b*). Meanwhile, the *offcut direction* is described by the azimuth angle $\Phi_{\text{off}}$ in Fig. 2(*b*). Note that the offcut can dramatically change the diffraction properties. For example, Fig. 2(*c*) shows the diffraction geometry for $\Phi_{\text{off}} = 180°$ and $\Omega_{\text{off}} = 90°$, which is a symmetric Laue case for 2BD. When $\Phi_{\text{off}} \neq n \cdot 180°$ and $\Omega_{\text{off}} \neq 0$, the configuration corresponds to non-coplanar two-beam diffraction (2BD) if only **G** is activated (Huang *et al.*, 2025). Non-coplanar grazing-incidence 2BD is a typical example of this category that has important applications for studying in-plane (epitaxial) crystal surface/interface strains and structures (Stepanov *et al.*, 1998).

Afterwards, the user can add a third Bragg reflection to the reflection list, and the MBD program automatically calculates its azimuthal angle(s) $\Phi_{3\text{BD}}$ (see Fig. 1). When the principal azimuth angle is set to $\Phi_{3\text{BD}}$, it becomes a 3BD case, and the users may calculate the 2D intensity contour around $(\Phi_{3\text{BD}}, \theta_B)$ in the $\Phi$-$\theta$ space ($\theta_B$ the Bragg angle of **G**). The MBD program can search for other reflections that may also (nearly) satisfy their Bragg conditions around the principal azimuth angle. For example, for **G** = 004, reflections 111 and 113 of silicon always have exactly the same azimuth angle, indicating that this is a 000-004-111-113 4BD process. Then the user can freely add additional reflections to the reflection list. If the Bragg condition of any added reflection is not fulfilled, it does not affect the computation results (although it may slightly slow down the computation). The MBD program has no restriction about the number of reflections included. As shown in Fig. 1, the overall MBD interface for user input is very simple, but the program can perform powerful computations behind the scenes. In fact, the FCWDT is the ultimate method capable of treating all dynamical diffraction cases in 3D space (including the above 2D coplanar two-beam Bragg and Laue cases) with a universal $4N \times 4N$ matrix method, where $N$ is the number of participating reflections (including $\mathbf{G}_0 = 000$).

Fig. 3(*a*) shows the reflectivity contour of the primary reflection Si 004 at $E$ = 8 keV under the 004-111-113 4BD conditions. Without the influence of reflections 111 and 113, the 004 reflectivity contour is a horizontal band independent of $\Phi$. Around the 4BD angular range, the secondary reflections 111 and 113 share the incident energy, leading to "loss of intensities" and distortions of the diffraction pattern for the primary reflection $\mathbf{G}$.

Note that a 3BD process involving reflections $\mathbf{G}_0$ (= 000), $\mathbf{G}_1$ (primary reflection) and $\mathbf{G}_2$ (operative reflection) has a hidden reflection channel $\mathbf{G}' = \mathbf{G}_1 - \mathbf{G}_2$, called the *cooperative reflection* by Cole *et al.* (1962). For 3BD to occur, only two of the three reflections $\mathbf{G}_1$, $\mathbf{G}_2$ and $\mathbf{G}'$ are required to be *allowed reflections*. This means that, for the cubic diamond structure, the triplet (000, 111, 002), as an example, is an allowed 3BD combination although 002 is a forbidden reflection. Here X-rays can be diffracted first by reflection 111 and then by the cooperative reflection $\mathbf{G}' = \mathbf{G}_1 - \mathbf{G}_2 = 11\bar{1}$ to form the 002 reflection. i.e., 002 is a "detour reflection" (Morelhão & Abramof, 1999; Huang *et al.*, 2014). Fig. 3(*b*) shows the intensity contour of the primary reflection $\mathbf{G}_1 = 111$ of Si under the 000-111-002 3BD conditions at $E$ = 8 keV. Compared with 111 two-beam diffraction, it is obvious that here the 111 reflection has the MBD features of intensity loss and diffraction pattern distortions. Note that the 000-111-002 case has a different reflection sequence than the 000-111-11$\bar{1}$ case although they involve the same reflections. The latter has a different azimuth angle $\Phi = -12.016°$ and the intensity contour of the primary reflection is a mirror image of the contour in the former case.

As another example, Fig. 3(*c*) shows the intensity contour of the primary reflection 002 in the 000-002-111 3BD process at $E$ = 8 keV. Here 002 itself is a forbidden direction, but the contour clearly shows its strong diffraction intensities around the 3BD center. These intensities are transferred by the detour reflection route 000 → 111 → $\bar{1}\bar{1}1$ (Lang *et al.*, 2013; Tang *et al.*, 2021; Zaumseil, 2015). When $\Phi$ is far away from $\Phi_{3BD}$, the 002 reflection tends to disappear. This is different from Figs. 3(*a*) and 3(*b*), where the primary reflection approaches two-beam diffraction when $\Phi$ is far away from $\Phi_{3BD}$. Here the simulation in Fig. 3(*a*) is consistent with the simulation by Estradiote *et al.* (2025) using a different MBD program. Fig. 3(*d*) shows the convoluted intensity contour of the 002 reflection based on a real incident beam with divergence of 10 µrad along both of the two orthogonal directions perpendicular to $\mathbf{K}_0$ in Fig. 2(*a*). Meanwhile, the beam has a Gaussian spectrum with a FWHM bandwidth of 1 eV. It is obvious that these real beam conditions smooth the sharp features in Fig. 3(*c*).

Overall, our numerous tests verify that the MBD program works correctly for all diffraction configurations. The program can also calculate convoluted MBD intensity contours or curves for incident X-ray beams with 2D angular divergence and finite bandwidths.

In addition to the MBD search function, *DXRD* also provides an independent mapping program for drawing MBD lines [also called monochromator glitches (Bunker, 2010)] in the azimuth-energy ($\Phi$-$E$) space. In the literature, a commonly overlooked situation in mapping of MBD lines for Si, Ge, and diamond is that forbidden reflections as operative reflections were ignored. For example, for Si (111) monochromators, attention has been paid on glitches caused only by allowed reflections, such as 400, 220, 311, etc. When acting as operative reflections, however, forbidden reflections such as 200, 240, $22\bar{2}$, etc., can all produce glitches [see Fig. 3(b)] because the corresponding cooperative reflections ($\bar{1}11$, $\bar{1}31$, $\bar{1}13$) are allowed reflections. These reflections have been largely ignored in the past. *DXRD* fully takes into account the detour reflections. For Si, Ge and diamond crystals, *DXRD* uses the simple diamond structure rule to judge whether a reflection $hkl$ itself is allowed [$h, k, l$ are all even numbers with $h + k + l = 4n$ ($n$ any integer), or $h, k, l$ are all odd numbers] or forbidden. For general crystals, the program calculates the structure factor of each reflection. If the structure factor is below the threshold value (which the user can set), the reflection is considered a forbidden reflection. However, the forbidden reflection is not necessarily excluded unless the involved triplet ($\mathbf{G}_1$, $\mathbf{G}_2$ and $\mathbf{G}' = \mathbf{G}_1 - \mathbf{G}_2$) has less than two allowed reflections. Calculations of the structure factors for thousands of reflections used to be time-consuming, but *DXRD* again uses a finely tuned algorithm that makes the calculations and mapping very fast (within a few seconds for most cases).

Another phenomenon ignored in the literature is that vertical MBD lines are usually missing in $\Phi$-$E$ maps. Such lines correspond to the continuous MBD geometry described by Huang *et al*. (2014). A typical example is Si 004 reflection with the HRD along [100], which is always a 000-004-022-$0\bar{2}2$ 4BD case at $\Phi = 0$ for any photon energy. Therefore, it corresponds to a vertical line at $\Phi = 0$ in the 004 reflection map of Fig. 4. However, this continuous MBD line seems missing in the map computed by the *MARE* program of *XOP*. Note that all allowed reflections of Si except 111 and 220 have vertical continuous MBD lines. In designing X-ray monochromators and analyzers, it is critical to avoid these lines because they always activate MBD at any diffraction energies. *DXRD* can reliably reveal such lines. Overall, the $\Phi$-$E$ mapping program of *DXRD* is useful for supporting MBD computations and for designing monochromators and analyzers. It is also helpful for interpreting X-ray spectroscopy patterns. It is worth noting that Rossmanith (2003) has made important achievements on the calculation and graphical representation of MBD patterns, particularly with the *UMWEG* program. This program has a unique function that can quickly calculate the *approximate intensity profile* of the glitches (which has also been implemented in *XOP*), but this function is not implemented in *DXRD*. Although the MBD program can calculate the accurate (convoluted) intensities of glitches [e.g., Fig. 3(d)] based on the rigorous dynamical theory, the computation is much slower.

## 4. Summary

We have introduced a new software tool, *DXRD*, for computing almost all kinds of X-ray diffraction from perfect single crystals using the dynamical theory. It has straightforward GUIs that allow general users to run the programs without the necessities of in-depth dynamical theory knowledge. The interactive input interfaces always automatically check the validity of the input parameters and guide the user to run the program. *DXRD* can calculate the conventional two-beam diffraction rocking curves of both the Bragg and Laue cases (including grazing-incidence geometry and backward diffraction) for both plane waves and divergent beams with finite spectral bandwidths. In particular, MBD computations have been extremely difficult even for experts in the past. *DXRD* provides an ultimate program that has a universal and fast algorithm, with which the users can easily perform complicated MBD computations. Since it treats X-ray diffraction in 3D space using a full vectorial matrix method, this algorithm is rigorous and applicable for all kinds of *N*-beam diffraction cases in 3D space (including coplanar and non-coplanar two-beam diffraction). *DXRD* also provides a mapping tool for accurately drawing all MBD lines in the azimuth-energy space, and this function is useful for indexing monochromator glitches in X-ray spectroscopy-related fields.

## Acknowledgements

We would like to thank Quanjie Jia and Zheng Tang for many helpful discussions and for debugging the programs. This work was supported by the U.S. DOE Office of Science-Basic Energy Sciences, under Contract No. DE-AC02-06CH11357.

# References

Authier, A. (2001). *Dynamical Theory of X-ray Diffraction. International Union of Crystallography Monographs on Crystallography*, No. 11. IUCr/Oxford University Press.

Bartels, W. J. (1983). *J. Vac. Sci. Technol. B* **1**, 338–345.

Brennan, S. & Cowan, P. L. (1992). *Rev. Sci. Instrum.* **63**, 850-853.

Bunker, G. (2010). *Introduction to XAFS: A Practical Guide to X-ray Absorption Fine Structure Spectroscopy*, Cambridge: Cambridge Univ. Press.

Chang, S. L. (2004). *X-ray Multiple-Wave Diffraction: Theory and Applications*, Solid State Sciences Series, Vol. 143. Berlin: Springer-Verlag.

Cole, H., Chambers, F. W. & Dunn, H. M. (1962). *Acta Cryst.* **15**, 138-144.

Colella, R. (1974). *Acta Cryst. A***30**, 413–423.

Estradiote, M. B., Nisbet, A. G. A., Penacchio, R. F. S., Miranda, M. A. R., Calligaris, G. A. & Morelhão, S. L. (2025). *J. Appl. Cryst.* **58**, 859-868.

Huang, X., Assoufid, L. & Macrander, A. L. (2025). *J. Synchrotron Rad.* **32**, 90-95

Huang, X.-R., Jia, Q., Wieczorek, M. & Assoufid, L. (2014). *J. Appl. Cryst.* **47**, 1716-1721.

Huang, X.-H., Peng, R.-W. & Fan, R.-H. (2010). *Phys. Rev. Lett.* **105**, 243901.

Huang, X.-H., Peng, R.-W., Hönnicke, M. G. & Gog, T. (2013). *Phys. Rev. A* **87**, 063828.

Lang, R., de Menezes, A. S., dos Santos, A. O., Reboh, S., Meneses, E. A., Amaral, L. & Cardoso, L. (2013). *J. Appl. Cryst.* **46**, 1796–1804.

Loxley, N., Tanner, B. K. & Bowen, D. K. (1995). *J. Appl. Cryst.* **28**, 314-317.

Morelhão, S. L. & Abramof, E. (1999). *J. Appl. Cryst.* **32**, 871-877.

Rossmanith, E (2003). *J. Appl. Cryst.* **36**, 1467-1474.

Sánchez del Río & M., Dejus, R. J. (2011). *Proc. SPIE* **8141**, *Advances in Computational Methods for X-Ray Optics II*, 814115. (https://doi.org/10.1117/12.893911)

Shvyd'ko, Y. (2004). *X-Ray Optics, High-Energy-Resolution Applications*, Berline: Springer.

Stepanov, S. (2021). *J. Appl. Cryst*. **54**, 1530-1534.

Stepanov, S. A., Kondrashkina, E. A., Koehler, R., Novikov, D. V., Materlik, G. & Durbin, S. M. (1998). *Phys. Rev. B* **57**, 4829-4841.

Stetsko, Y. P. & Chang, S.-L. (1997). *Acta Cryst. A* **53**, 28-34.

Tang, Z., Zheng, L., Chu, S., An, P., Huang, X., Hu, T. & Assoufid, L. (2021). *J. Appl. Cryst*. **54**, 976-981.

Zaumseil, P. (2015). *J. Appl. Cryst.* **48**, 528-532.

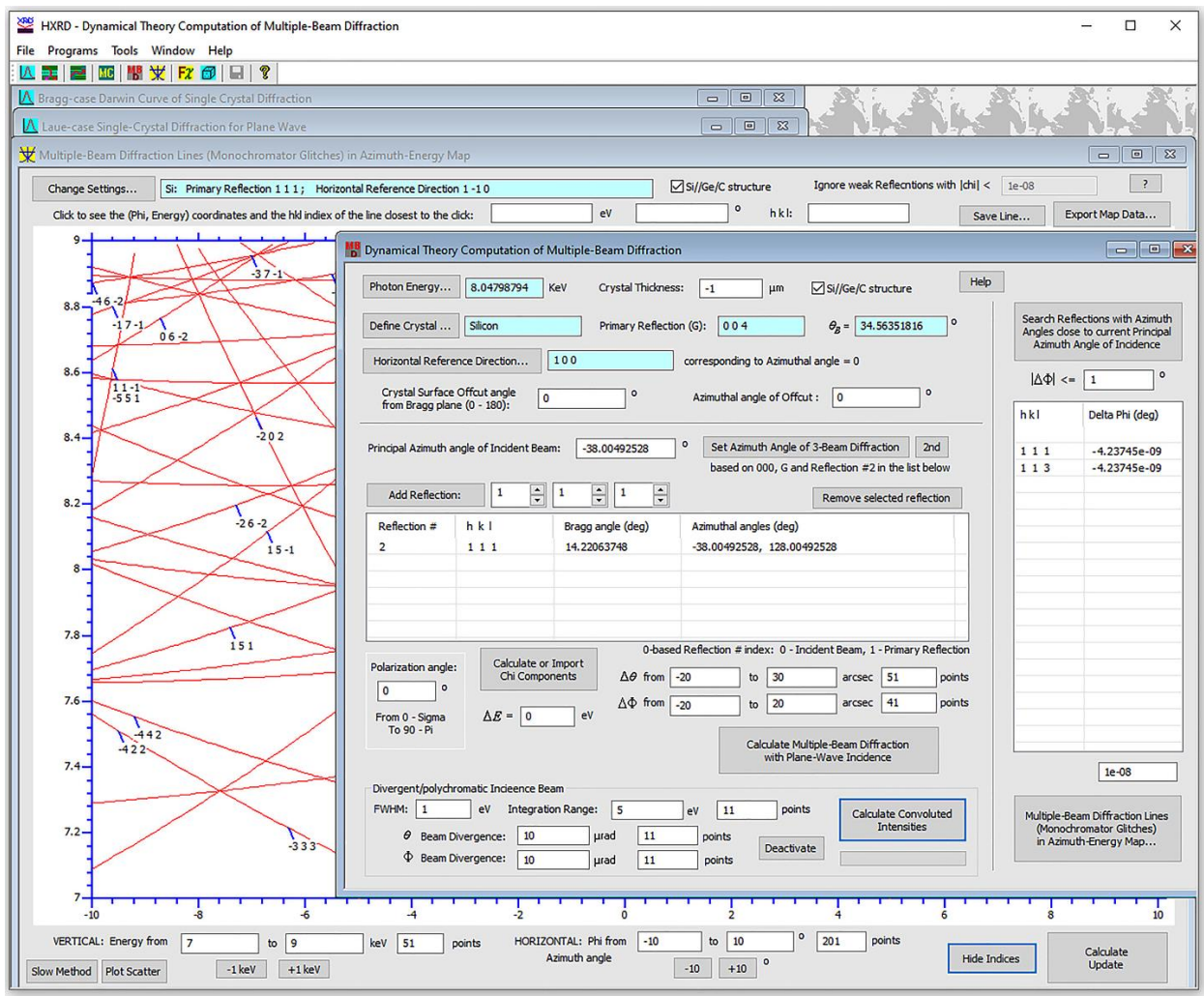

Fig. 1 (Please make it two column wide if possible) The interfaces of *DXRD*. All the programs are integrated into a single app. The front window is the interface of the MBD program. The typical workflow for most of the programs includes the simple step sequence of (1) defining the X-ray energy, (2) importing the crystal structure, (3) selecting the reflection(s), (4) setting the crystal thickness, (5) specifying surface offcut, (6) setting calculation range(s), and (7) performing the calculations.

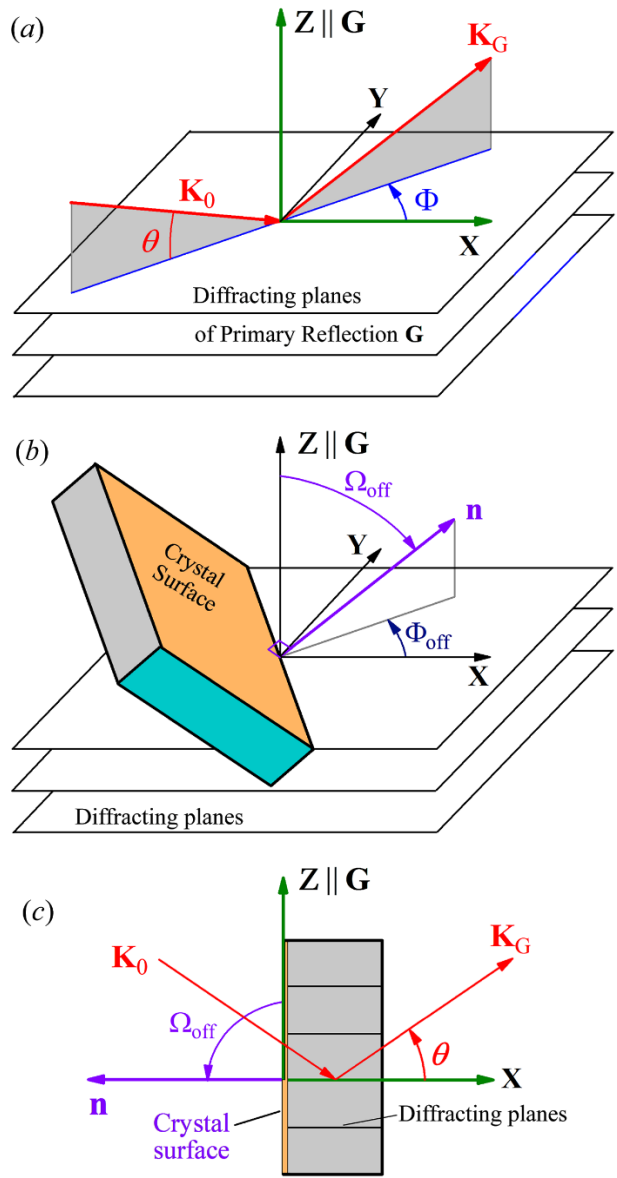

Fig. 2. The CS used by the MBD program for all diffraction configurations. (*a*) The CS related to the primary reflection **G**. (*b*) The parameters used for describing the offcut of the crystal surface with respect to the **G** diffracting planes. (*c*) The surface offcut parameters for symmetric Laue transmission of **G**.

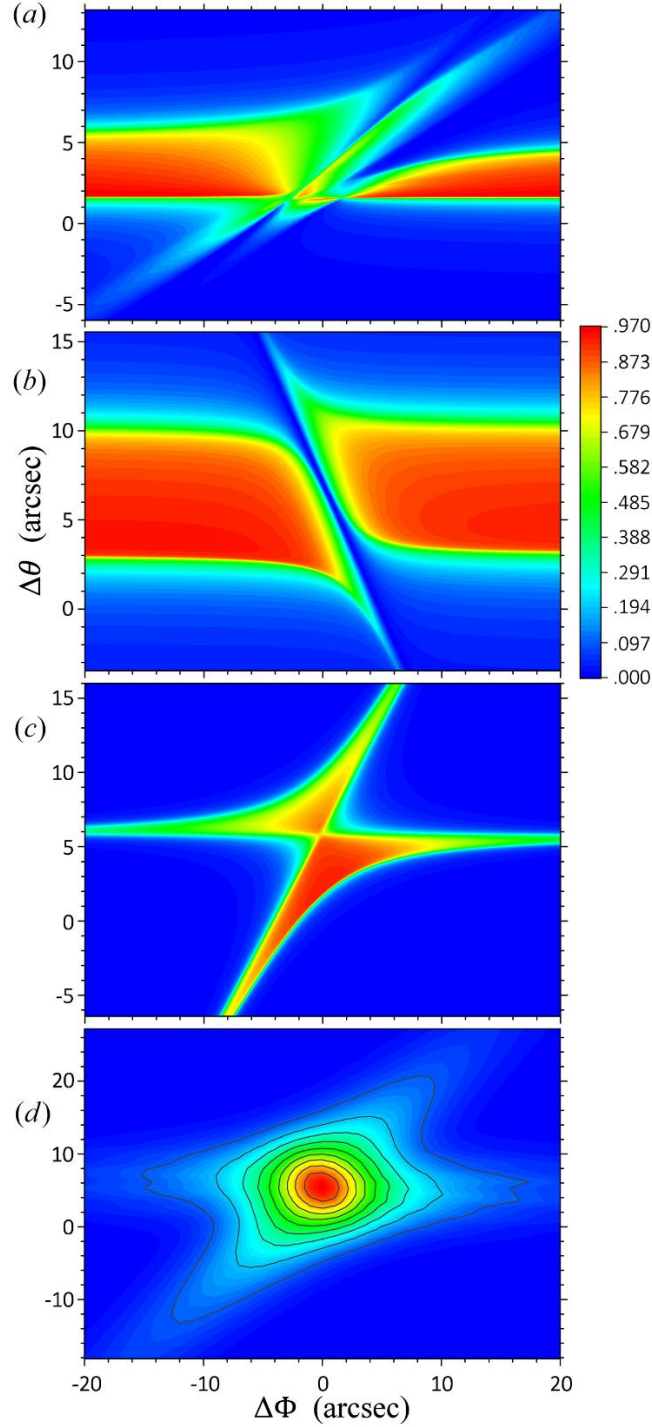

Fig. 3. Examples of MBD reflectivity contours computed by *DXRD*. (*a*) Reflectivity contour of the primary reflection 004 in 000-004-111-113 4BD with the HRD [100]. The principal azimuth angle is $\Phi_{4BD} = -37.94°$, where the 4BD condition is fulfilled. $\Delta\Phi = \Phi - \Phi_{4BD}$. $\Delta\theta = \theta - \theta_B$. (*b*) 111 reflectivity contour in 000-111-002 3BD. $\Phi_{3BD} = 16.58°$ from the HRD $[1\bar{1}0]$. (*c*) Reflectivity contour of the forbidden primary reflection 002 in 000-002-111 3BD. $\Phi_{3BD} = -51.04°$ from the HRD [100]. The incident beam is a plane wave with $E = 8$ keV for (*a*)-(*c*). (*d*)

Intensity contour of reflection 002 under the same conditions of (c) except that the incident beam has angular divergence of 10 μrad along both the $\theta$ and $\Phi$ directions and a spectral bandwidth of 1 eV. σ-polarization. $\Phi_{off} = \Omega_{off} = 0$.

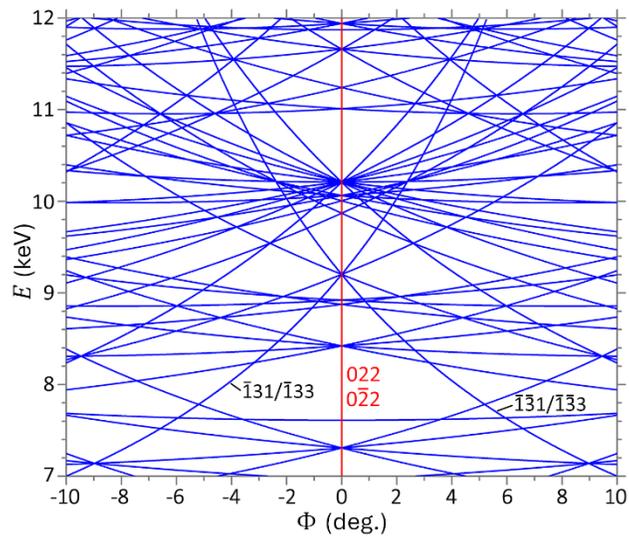

Fig. 4. MBD lines in the $\Phi$-$E$ map of primary reflection Si 004. The red line corresponds to continuous 000-004-022-0$\bar{2}$2 4BD that always occurs for any energy $E$ at $\Phi = 0$.